\documentstyle[prl,aps,preprint,mypsfig2,floats]{revtex}

\def\@pnumwidth{2em}

\makeatother

\begin{document}
\noindent {\Large \sf Nuclear Magnetic Resonance Quantum Computing \\
     Using Liquid Crystal Solvents}\\

\noindent{\large \sf Costantino S. Yannoni, Mark H. Sherwood, Dolores C. Miller, \\and Isaac L. Chuang}\\
IBM Almaden Research Center, San Jose, CA 95120\\
{\large \sf Lieven M.K. Vandersypen}\\
Solid State  and Photonics Laboratory, Stanford University,  Stanford, 
CA 94305-4075, and IBM Almaden Research Center, San Jose, CA 95120\\
{\large \sf Mark G. Kubinec}\\
College of Chemistry, D62H Hildebrand Hall, University of California Berkeley, Berkeley, CA 94720-1460\\

\date{\today}


\def\>{\rangle}
\def\be{\begin{equation}}
\def\ee{\end{equation}}
\def\bea{\begin{eqnarray}}
\def\eea{\end{eqnarray}}
\newcommand{\ket}[1]{\mbox{$|#1\rangle$}}
\newcommand{\bra}[1]{\mbox{$\langle #1|$}}
\newcommand{\mypsfig}[2]{\psfig{file=#1,#2}}



\vspace*{-0.5cm}
\begin{abstract}
\vspace*{-1cm}

Liquid crystals offer several advantages as solvents for molecules used for
nuclear magnetic resonance quantum computing (NMRQC).  
The dipolar coupling between nuclear spins
manifest in the NMR spectra of molecules oriented by a liquid crystal permits
a significant increase in clock frequency, while short spin-lattice relaxation
times permit fast recycling of algorithms, and save time in calibration and
signal-enhancement experiments.  Furthermore, the use of liquid crystal
solvents offers scalability in the form of an expanded library of spin-bearing
molecules suitable for NMRQC.  These ideas are demonstrated with the
successful execution of a 2-qubit Grover search using a molecule
($^{13}$C$^{1}$HCl$_3$) oriented in a liquid crystal and a clock speed eight
times greater than in an isotropic solvent.  Perhaps more importantly, five
times as many logic operations can be executed within the coherence time using
the liquid crystal solvent.
\end{abstract}




Quantum computations have been done using NMR techniques to manipulate
ensembles of coupled nuclear spins (qubits) in molecules in
solution.\cite{nmrqc} Isotropic solvents are favorable for NMRQC experiments
since the narrow, well-resolved NMR lines satisfy the dual requirements of
individual spin addressability and long coherence time.\cite{divincenzo}
However, the quantum computer clock is slow, since the time required to
execute logic gates is approximately $1/2J$ seconds ~\cite{cg} where $J$, the strength
of the spin-spin scalar coupling that persists in isotropic solution, is
typically less than $300$ Hz for organic solutes.\cite{scalar}  In
liquid
crystal solvents, however, the solute molecules become partially oriented
against a background of rapid Brownian motion, resulting in well-resolved NMR
spectra characterized by dipolar coupling between nuclear spins, which
can be as large as $10$ kHz.\cite{emsley} Thus, the NMRQC clock in liquid
crystal solvents can be much faster.  Several other advantages of using liquid
crystals for NMRQC experiments are outlined below.


The behavior of $N$ coupled spins in a molecule dissolved in an isotropic liquid
in
an external magnetic field is given by the Hamiltonian
\be
     H_{\rm iso}/h = \sum_{i}^N \nu_i I_{zi} + \sum_{i<j}^N J_{ij} (I_{xi}
I_{xj} +
I_{yi} I_{yj} + I_{zi} I_{zj})
\label{eq:isoham}
\ee
where $\nu_i$ is the resonance frequency (Hz), $I_{zi}$ is the z component of
the angular momentum operator, and $J_{ij}$ is
the strength of the scalar coupling (Hz).\cite{abragam}  For two spins A and B,
which yield a first order NMR spectrum, i.e. $|\nu_A - \nu_B| >> |J|$,
Eq.(\ref{eq:isoham}) becomes )\cite{abragam}
\be
     H_{ iso}^{\circ} /h = \nu_A I_{zA} + \nu_B I_{zB} + J I_{zA} I_{zB}.
\label{eq:isoham0}
\ee
The time evolution of spins under $H_{ liq}^{\circ}$
permits the use of relatively simple pulse sequences to do NMR
quantum computing.\cite{cg}  The clock frequency, $f_{clock}$, for an NMR
quantum computer based on a
two-spin system with a scalar coupling of 200 Hz is given by 2$|J|$ $\approx$ 400 Hz.

The Hamiltonian for $N$ spins in a molecule dissolved in a liquid
crystal solvent is
\be
     H_{\rm lc}/h
     = \sum_{i}^N \nu'_i I_{zi} + \sum_{i<j}^N J_{ij} (I_{xi} I_{xj} + I_{yi}
I_{yj}
+ I_{zi} I_{zj})
          + \sum_{i<j}^N D_{ij} \left\{ { 2 I_{zi} I_{zj}
               - \frac{1}{2} (I_{xi} I_{xj} + I_{yi} I_{yj})
               }\right\}.
\label{eq:lxtal}
\ee
A dipolar term now appears because the molecules are partially oriented and
the resonance frequency $\nu'_i$ includes the effects of
molecular orientation and chemical shift anisotropy.\cite{emsley}  The dipolar
coupling
strength $D$, which also depends on the orientation, is typically $100$ Hz to
$10$ kHz.  However, the pulse sequences used for NMRQC in isotropic liquids
can not be applied because of the form of the spin operators in
Eq.(\ref{eq:lxtal}).\cite{warren}  For a 2-spin system with a first order
spectrum,  Eq.(\ref{eq:lxtal}) becomes\cite{saupe}
\be
     H_{\rm lc}^{\circ}/h
     = \nu'_A I_{zA} + \nu'_B I_{zB} + (J+2D) I_{zA} I_{zB}.
\label{eq:lxtaltwo}
\ee
Since Eq.(\ref{eq:lxtaltwo}) has the same form as Eq.(\ref{eq:isoham0}), the
pulse sequences that have been used successfully for NMRQC in isotropic
solution\cite{nmrqc} can be applied directly to liquid crystal solutions,
permitting computations with $f_{clock} = 2|(J + 2D)|$ Hz, a frequency that can
be much higher than $2|J|$.


This is borne out in Table~\ref{tab:relax} which shows the $^{13}$C - $^{1}$H
coupling strength, spin-lattice relaxation time ($T_1$) and spin-spin
relaxation (decoherence) time ($T_2$) for $^{13}$C and $^{1}$H in chloroform
($^{13}$C$^{1}$HCl$_3$) in both liquid crystal and isotropic solution at
ambient temperature.\cite{expts}

The $^{13}$C-$^{1}$H coupling in the liquid crystal (ZLI-1167\cite{lxtal}) is
eight times larger than the scalar coupling in acetone-d$_6$, corresponding to
a computer with a clock that is eight times faster.  The product of the
shortest coherence time and the clock frequency $T_2 f_{clock} = 2T_2J$, which
approximates the number of gates that can be executed while maintaining
coherence,
may be used as a figure of merit.  The data in Table~\ref{tab:relax} show that
$T^{lc}_{2}f^{lc}_{clock} \approx 5 T^{iso}_{2}f^{iso}_{clock}$, meaning that
more complex algorithms requiring five times as many logic operations can be
executed using this solute/liquid-crystal-solvent system.  The chloroform
$^{13}$C and $^1$H
spin-lattice relaxation times are about $12$ times shorter in ZLI-1167 than in
acetone-$d_6$.  Since all NMRQC algorithms as well as NMR experiments used to
set up and calibrate the spectrometer require a polarization time of $5T_1$ s,
an order of magnitude savings in time can be significant, and will become more
so
as the number of qubits increases.  This advantage will also be manifest when
the sensitivity must be increased by co-addition of the signal from several
successive NMRQC experiments, or when procedures which require multiple
experiments such as quantum state tomography\cite{tomography} are used to
diagnose the operation of quantum algorithms.

Another advantage of using liquid crystals as solvents for NMRQC is that they
permit a wider choice of  spin-bearing molecules that may be suitable for NMRQC.
Dipolar coupling, which is manifest in the NMR spectra of
oriented molecules, requires only proximity between the spins of interest.  As
a result, two spins that are separated by several bonds and which have no
scalar coupling may, if spatially proximate, have dipolar coupling
sufficiently large for quantum computation.  The ability to control the degree
of orientation of the solute molecule by varying the solvent temperature and
solute concentration \cite{emsley} provides the experimentalist with means of
tailoring the NMR spectrum to meet the requirements for NMRQC.  In addition,
magic-angle spinning and multiple pulse methods can be used to preferentially
scale the dipolar
splitting in the spectrum of a liquid-crystal-oriented molecule to convert it
to first order.\cite{cortieu}

Complications do arise with the use of liquid crystal solvents: (1) the NMR
lines of small molecules dissolved in liquid crystal
solvents are susceptible to a broadening mechanism not found in isotropic
solution, most likely due to variations in the degree of
orientation caused by thermal gradients in the sample.  Nonetheless,
resonance line widths $<2$ Hz ($^{13}$C) and $<3$ Hz ($^{1}$H) were obtained
for $^{13}$C$^{1}$HCl$_3$ in ZLI-1167; (2) the large dipolar couplings may 
cause unwanted evolution of the spins during the relatively long pulses
required for selective excitation in homonuclear spin systems.


In order to show that quantum computations can be done successfully using
liquid-crystal solution NMR, we have implemented the Grover search
algorithm\cite{grover} using $^{13}$C$^{1}$HCl$_3$ dissolved in ZLI-1167.  
The goal of the search is as follows: given a function
$f(x)$, find the unknown element $x_0$ among four possible elements
00,01,10,11 - represented by the four spin-product states
$\ket{00},\ket{01},\ket{10},\ket{11}$ - which satisfies $f(x_0) = 1$, where
$f(x) = 0$ for the other three elements.  Classically, this would take an
average of $2.25$ attempts, while one query is sufficient using the Grover
algorithm.  
The carbon and proton spins were first prepared in an effective pure state 
created by temporal labeling\cite{knill} followed by a previously used Grover
protocol.\cite{lieven}
The prediction is that the algorithm will put the spins in the
state $\ket{x_0}$.  The $^{13}$C and $^1$H readout spectra for the four
possible $x_0$ are shown in Fig.~\ref{fig:grover}.  As predicted for two spins
in an effective pure state, the value of $x_0$ is clearly indicated by the
amplitude and phase of the
two resonance lines in the $^{13}$C and $^1$H spectra.  Measurement of the
deviation density matrix using quantum state tomography\cite{tomography},
confirms the output states to be as theoretically predicted; 
Fig.~\ref{fig:denmat} shows that for $x_0=11$, the final state of the spins 
is the state $|11\>$.

An immediate advantage of using liquid crystal solvents becomes clear from
the savings in experimental time.  A wait-time ($5T_1$) between experiments of
only $8$ s sufficed, compared with $105$ s in acetone-$d_6$.\cite{nmrqc} The
increase in spin coupling strength more than compensated for the shorter
coherence time and permitted successful completion of the algorithm.


The use of liquid crystals as solvents in NMR quantum computing has been
proposed and demonstrated.  Due to the large $^{13}$C-$^{1}$H dipolar
coupling, the speed of the NMR quantum computer was increased by a factor of
$8$ while the short spin-lattice relaxation times resulted in significant time
saved in setting up the spectrometer and in executing the Grover search
algorithm.  This demonstration of the utility of liquid crystals as NMRQC
solvents expands the library of potential quantum computing molecules.


The authors would like to acknowledge helpful discussions on liquid crystal
solvents with B. M. Fung.  The support and encouragement of Nabil Amer,
James Harris and Alex Pines are also appreciated. L.V. Acknowledges a
Yansouni Family Stanford Graduate Fellowship. This work was performed
under the auspices of the DARPA NMRQC initiative.

\clearpage

\begin{center}
\begin{table}[htbp]
\begin{center}
\caption{$^{13}$C-$^{1}$H spin coupling and relaxation times for $^{13}$C$^{1}$HCl$_3$ in isotropic and liquid crystal (ZLI-1167) solution. Times are in seconds and couplings in Hz.}
\label{tab:relax}
\vspace*{1cm}
\begin{tabular}{ccccccc}
solvent & $J$ & $J+2D$ & $T_1$ ($^{13}$C)   &  $T_1$ ($^{1}$H)
     & $T_2$ ($^{13}$C) &  $T_2$ ($^{1}$H)
\\\hline
acetone-$d_6$ & 215 & \rule{2ex}{0.2pt} & 25 & 19 & 0.3 & 7 \\
ZLI-1167 & \rule{2ex}{0.2pt} & 1706 & 2 & 1.4 & 0.2 & 0.7 \\
\end{tabular}
\end{center}
\vspace*{0.5cm}
\end{table}
\end{center}

\clearpage

\begin{center}
{\large FIGURE CAPTIONS}
\end{center}

\begin{figure}[htbp]
\caption{Spectral readout of the results of the 2-qubit Grover search
using $^{13}$C$^{1}$HCl$_3$ in a liquid crystal solvent showing absorption and
emission
peaks which clearly indicate $x_0$ equal to $00$, $01$, $10$, and $11$
(from top to bottom).  The real part of the $^1$H (left) and $^{13}C$
(right) spectra are shown, with NMR lines at $\nu_H \pm J_{CH}/2$ and $\nu_C
\pm J_{CH}/2$ (shown in kHz relative to $\nu_H$ and $\nu_C$).  The
vertical scale is arbitrary.  }
\label{fig:grover}
\end{figure}

\vspace*{2cm}

\begin{figure}[htbp]
\caption{Experimentally measured deviation density matrix elements for the 11 case.}
\label{fig:denmat}
\end{figure}

\clearpage

\begin{center}
$$
\begin{array}{cc}
\mbox{\psfig{file=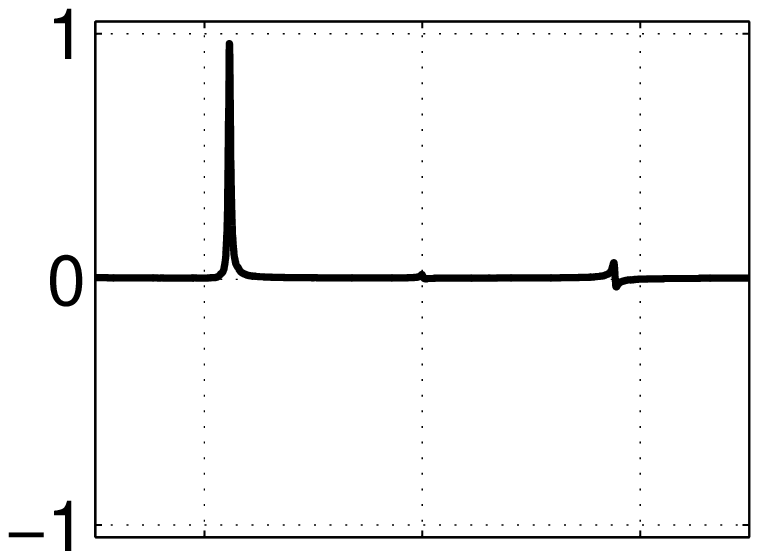,width=2in}} &
\mbox{\psfig{file=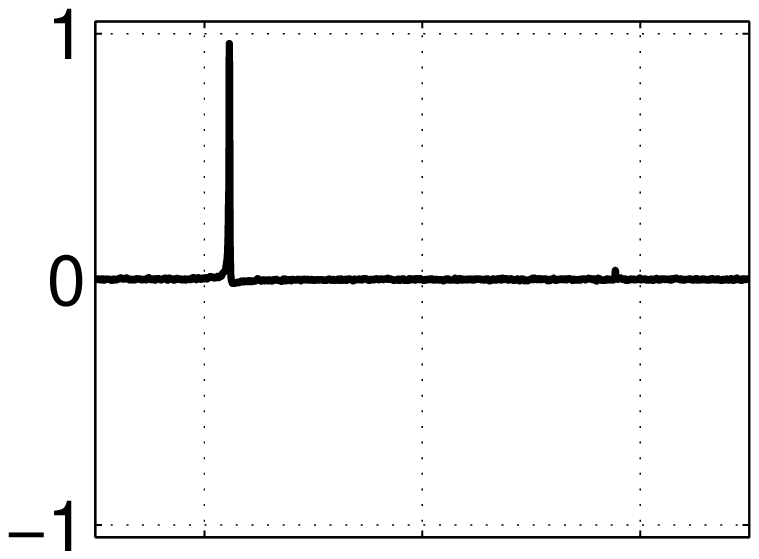,width=2in}} \\
\mbox{\psfig{file=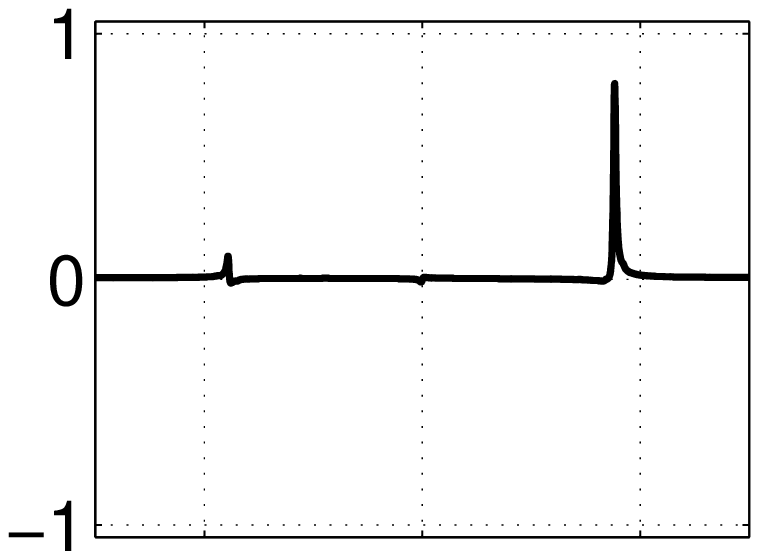,width=2in}} &
\mbox{\psfig{file=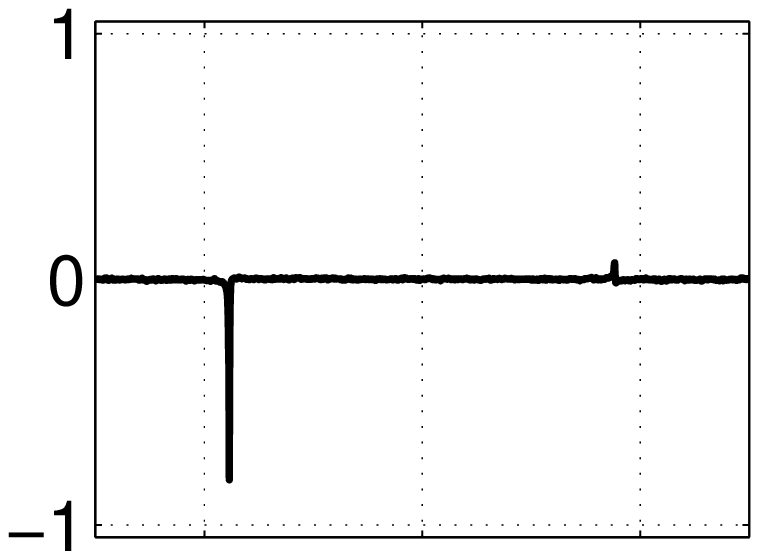,width=2in}} \\
\mbox{\psfig{file=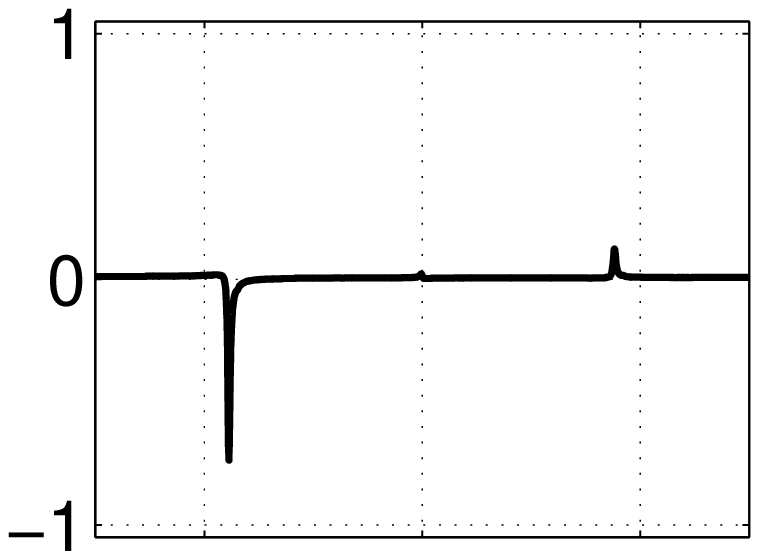,width=2in}} & 
\mbox{\psfig{file=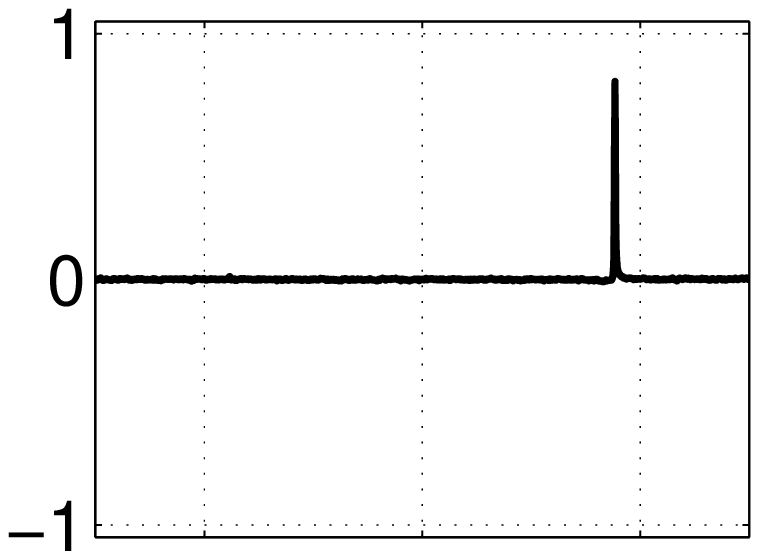,width=2in}} \\
\mbox{\psfig{file=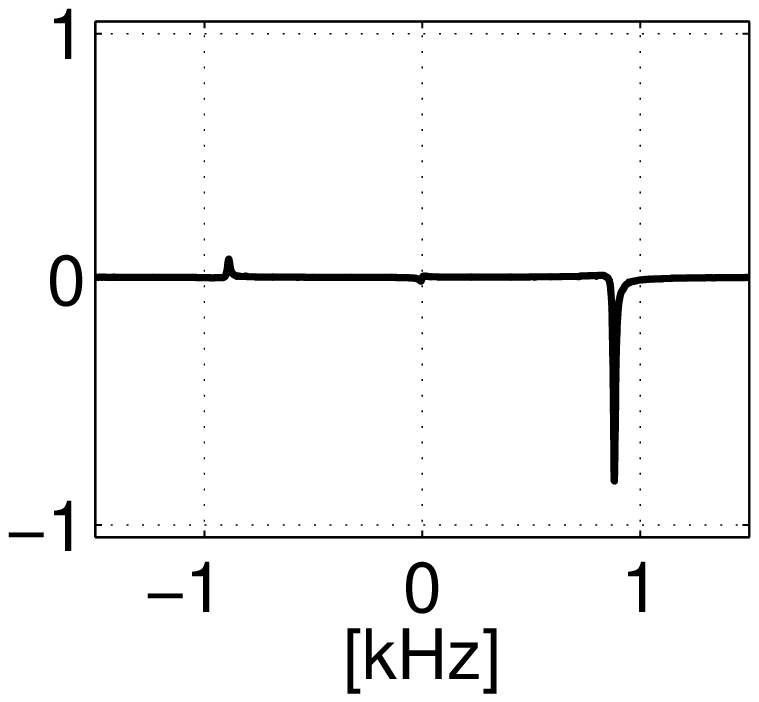,width=2in}} & 
\mbox{\psfig{file=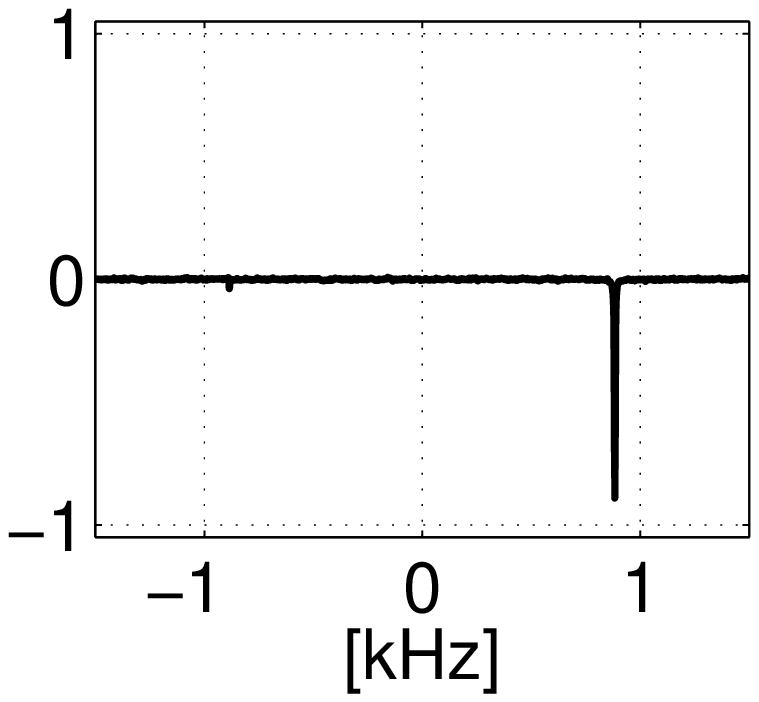,width=2in}} \\
\end{array}
$$
\end{center}
\vspace*{3cm}

Figure 1 - Costantino Yannoni, Applied Physics Letters

\clearpage

\begin{center}
\mbox{\psfig{file=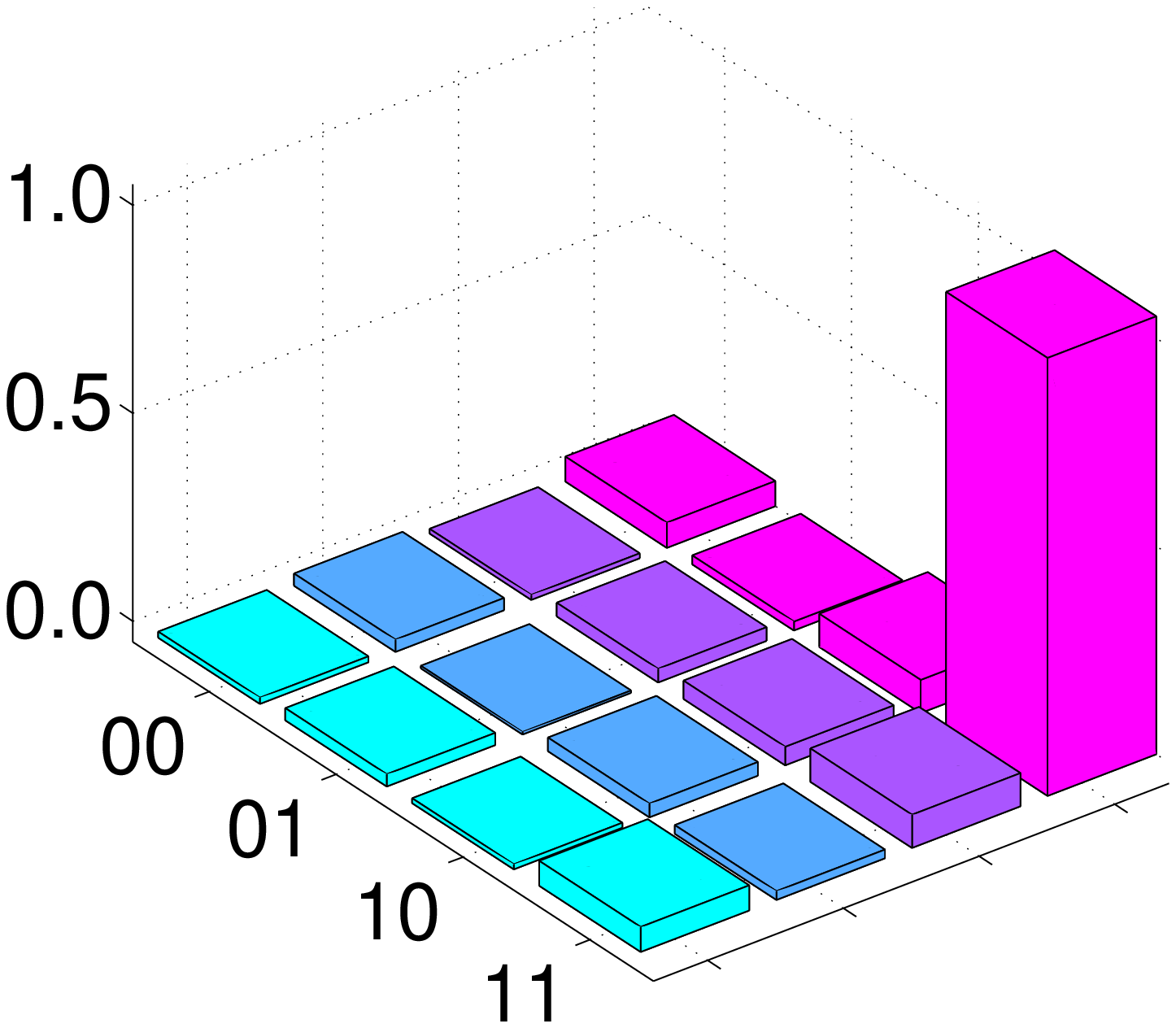,width=3.5in}}
\end{center}
\vspace*{6cm}
Figure 2 - Costantino Yannoni, Applied Physics Letters


\end{document}